\documentclass[11pt]{revtex4}

\usepackage{graphicx}
\usepackage{epstopdf}
\usepackage{caption}
\usepackage[latin1]{inputenc}
\usepackage{amsmath}
\usepackage{amsfonts}
\usepackage{amssymb}
\usepackage{makeidx}
\usepackage{color}
\usepackage{mathrsfs}
\usepackage{amsfonts}
\RequirePackage[colorlinks,citecolor=blue,urlcolor=magenta,linkcolor=blue]{hyperref}
\usepackage{multirow}
\usepackage{array}
\begin{document}

\title{Modified Bekenstein-Hawking system in $f(R)$ gravity}

\author{Jibitesh Dutta\footnote {jdutta29@gmail.com, jdutta@associates.iucaa.in}}
\affiliation{Mathematics Division, Department of Basic Science and
Social Sciences, North Eastern Hill University, Shillong 793022,
Meghalaya, (India)\\ and\\ Visiting Associate, Inter University
Centre for Astronomy and Astrophysics, Pune 411 007,  (India)}
\author{Saugata Mitra\footnote{saugatamitra20@gmail.com}}
\affiliation{Department of Commerce (Evening), St. Xavier's College (Autonomous), 30 Mother Teresa Sarani, Kolkata 700016, India.}
\author{Binod Chetry\footnote {binodchetry93@gmail.com}}

\affiliation{Department of Mathematics, North Eastern Hill University, Shillong 7930222, Meghalaya, India.}

\begin{abstract}
\noindent The present work deals with four alternative formulation
of Bekenstein system on event horizon in $f(R)$ gravity. While
thermodynamical laws holds in universe bounded by apparent
horizon, these laws break down on event horizon.  With alternative
formulation of thermodynamical parameters (temperature and
entropy), thermodynamical laws hold on event horizon in Einstein
Gravity. With this motivation, we extend the idea of generalised
Hawking temperature and modified Bekenstein entropy in homogeneous
and isotropic model of universe on event horizon  and examine
whether thermodynamical laws hold in  f(R) gravity. Specifically,
we examine and compare validity of generalised second law of
thermodynamics (GSLT) and thermodynamical equilibrium (TE) in four
alternative modified Bekenstein scenarios. As Dark energy is a
possible dominant candidate for matter in the univerese  and
Holographic Dark Energy (HDE) can give effective description of
f(R) gravity, so matter in the universe is taken as in the form
interacting HDE.  In order to understand the complicated
expressions, finally the above laws are examined from graphical
representation using three Planck data sets and it is found that
generalised/modified Hawking temperature
 has a crucial role in making perfect thermodynamical system.
\end{abstract}

\maketitle



\section{Introduction}
\noindent One of the interesting topic in general relativity (GR)
is the relation between thermodynamics and gravity. This relation
helps in understanding several aspects of GR
\cite{Padmanabhan:2013nxa}. The relations were first observed by
Hawking and Bekenstein in $1970$ in the context of black hole
thermodynamics and it was realised black hole can be considered as
thermodynamical system with a temperature and entropy.
\cite{Hawking:1974sw}.  The temperature and entropy are
proportional to the surface gravity and horizon area respectively
\cite{Hawking:1974sw,Bekenstein:1973ur}. Further, the first law of
thermodynamics relates these temperature and entropy
\cite{Bardeen:1973gs}. As entropy is a thermodynamical quantity
and horizon is a geometrical quantity, people started predicting
relation between black hole thermodynamics and Einstein field
equations. Indeed,  Jacobson in $1995$  derived the Einstein field
equations from the first law of thermodynamics. Padmanabhan, from
the other side was able to derive first law of thermodynamics for
a general static spherically symmetric space-time
\cite{Padmanabhan:2002sha, Padmanabhan:2003gd}.

\noindent Subsequently these ideas are generalised in cosmology,
treating universe as a thermodynamical system. Generally apparent
horizon $(R_A)$ is taken as boundary of the universe. The
thermodynamics in de-Sitter's space time was first investigated by
Gibbons and Hawking \cite{Gibbons:1977mu}. Furthermore, the first
law of thermodynamics and the Friedmann equations are shown to be
equivalent \cite{Cai:2005ra}. On the other hand from the present
observational data it is found that the current expansion of the
universe is accelerating \cite{Perlmutter:1998np, Riess:1998cb,
Astier:2005qq, Riess:2006fw, Spergel:2003cb, Peiris:2003ff,
Spergel:2006hy, Komatsu:2008hk}. The event horizon is distinct
from apparent horizon and its existence is assured in the
accelerating universe. Wang $et~al.$ \cite{Wang:2005pk} in $2006$
argued that the event horizon is larger than the apparent horizon
and the universe bounded by the event horizon is not a Bekenstein
system in the study of laws of thermodynamics when the universe is
having accelerated expansion. Further, they concluded that the
outside apparent horizon thermodynamical laws break down and
thermodynamic description does not follow Bekenstein's definition.

\noindent It is imperative to verify thermodynamical laws for
modified gravity if thermodynamic interpretation of gravity near
horizon is a generic feature. $f(R)$ gravity is one of the
prominent modified gravity which can explain  naturally the
present acceleration of the universe without any dark energy (DE).
In $f(R)$ gravity the action term is arbitrary function $f(R)$ of
the Ricci scalar $R$ . Some other classes of modified
gravities are $f(G)$, $f(R,G)$, and $f(T)$. These theories are
considered as gravitational alternatives for DE and extensively
explored for different purposes in the literature
\cite{Capozziello:2002rd, Nojiri:2006gh, Nojiri:2006be,
Nojiri:2009kx, Starobinsky:1980te, Kerner:1982yg, Barrow:1983rx,
Faraoni:2006hx, Schmidt:2006jt, Hu:2007nk,  Nojiri:2003ni,
Nojiri:2003wx, Abdalla:2004sw, Nojiri:2003rz, Carroll:2003wy,
Capozziello:2005mj, Bisabr:2008hy, Karami:2010aq,
KhodamMohammadi:2010py, Bengochea:2008gz, Karami:2013rda}.

\noindent In recent years, lot of work has been done in $f(R)$
gravity in the context of gravitational thermodynamics
\cite{Karami:2012hq, MohseniSadjadi:2007zq, Saiedi:2012qk,
Wu:2012ia, Ghosh:2012bd, Elizalde:2011zz, Yokokura:2011za,
Bamba:2009ay, Akbar:2006mq, Akbar:2006er}. Further, it has been
extended in other gravity theories including $f(G)$ theory
\cite{Cai:2005ra}, Scalar tensor gravity \cite{Akbar:2006er,
Akbar:2006mq, Karami:2014tsa}, Lovelock theory \cite{Akbar:2006kj}
and braneworld scenarios \cite{Sheykhi:2009zza}.  All these works
on universal thermodynamics mostly deal with apparent horizon but
in case of event horizon due to its complicated nature there are
few works related to it. It is imperative to investigate the
validity of thermodynamical laws for event horizon as it separates
out from apparent horizon in accelerating universe. In this regard
it has been proved that generalised second law of thermodynamics
holds in any gravity theory under some conditions in event horizon
\cite{Mazumder:2009zz, Mazumder:2009zza, Mazumder:2010fi,
Dutta:2010zza, Chakraborty:2010jx, Chakraborty:2010hr}. The main
difficulty of studying the thermodynamics of the universe is to
define the entropy and temperature on the horizons. Generally the
entropy and hence temperature is taken from black hole
thermodynamics but in modified gravity theories some corrections
term may be needed.

In literature various forms of entropy and temperature have been
proposed to study thermodynamics in expanding universe
\cite{Chakraborty:2012cw, Saha:2012nz, Chakraborty:2012eu}. For
instance by generalising the temperature or modifying the entropy
in Einstein gravity, validity of thermdynamical laws and TE have
been shown in various gravity theories
\cite{Mitra:2015wba,Mitra:2015nba, Mitra:2015jqa, Mitra:2015jha,
Saha:2015gha, Radicella:2010ss}. In other words, generalised
Hawking temperature or a modified  Bekenstein entropy on event
horizon helps in making perfect thermodynamical system in
different gravity theories. Therefore, it is natural to ask,
whether with these alternative definition of thermodynamical
parameters (temperature and entropy), thermodynamical laws hold on
event horizon in f(R) gravity.  With this motivation, here we
extend the idea of generalised Hawking temperature and modified
Bekenstein system on event horizon in $f(R)$ gravity. In this
regard, we study and compare four different types of modified
entropies/temperatures and examine the validity of thermodynamical
laws with respect to these modified entropies and temperatures.

Dark energy (DE) is a possible dominant candidate for matter in
the universe. One of the DE candidate which received lot of
attention in recent years is Holographic Dark Energy (HDE) as it
can alleviate coincidence problem \cite{Zimdahl:2007zz}. The HDE
model is based on the holographic principle which states that the
number of degrees of freedom of a physical system is given by the
area of the boundary \cite{'tHooft:1993gx, Susskind:1994vu}. This
model makes an attempt to apply holographic principle of quantum
gravity to DE problem and one can obtain HDE density as
 $\rho_D=3c^{2}M^2_{p}L^{-2}$, where
$M^2_p=8\pi G$ is the reduced Planck mass, $L$ is the IR (infra
red) cut-off (size of the region) and $c$ is numerical constant
\cite{Cohen:1998zx}. A comprehensive review of IR cut-off and
various cosmological implications of HDE in accelerating universe
can be obtained in Refs.\cite{Pavon:2006qm, mli, Pavon:2005yx,
Zimdahl:2007zz}. Further, HDE can give effective description of
f(R) gravity \cite{Wu:2007tn}. So in the present work matter in
the universe is taken as in the form of  interacting HDE. In order
to understand the complicated expressions, finally, we have
examined the validity of GSLT and TE from graphical representation
using three Planck data sets.

The outline of the paper is as follows. The section II presents
basic equations in $f(R)$ gravity. In section III we study the
basic concepts of gravitational thermodynamics. Section IV deals
with thermodynamical analysis in a universe dominated by HDE and
finally in last section we discuss the summary of the work and
possible conclusions.


\section{Basic equations of $f(R)$-gravity}

\noindent The modified Einstein-Hilbert action in $f(R)$ gravity is written as

\begin{equation}\label{eqn1}
S = \frac{1}{16\pi G}{\int {d^4}x\sqrt{-g}f(R)} + S_m,
\end{equation}
\noindent with $S_m$ as the matter action, $R$ is Ricci scalar and $f(R)$ is the arbitrary real function of $R$. Taking variation of the action (\ref{eqn1}) with respect to the metric $g_{\mu\nu}$, we get

\begin{equation}\label{eqn2}
R_{\mu\nu}f_R-\frac{1}{2}g_{\mu\nu}f-\nabla_{\mu}\nabla_{\nu}f_R+g_{\mu\nu}{\nabla}^{2}f_R=T^{m}_{\mu\nu},
\end{equation}
\noindent where $f_R$ denotes the derivative of $f$ with respect
to $R$ and $T^{m}_{\mu\nu}=\rm diag(-\rho, p, p, p)$ is the energy
momentum tensor for the matter field. The matter field is taken in
the form of a perfect fluid with $\rho=\rho_m +\rho_d$, $p=p_d$.
Here $\rho_d$ , $p_d$ respectively denote the energy density and
thermodynamic pressure of HDE and $\rho_m$ denotes the energy
density of matter (dark matter +Baryonic)

Observations support flat, homogeneous and isotropic FRW metric
given by

\begin{equation}\label{eqn3}
ds^2= h_{ij}(x^i)dx^i dx^j + R^2_h d{\Omega}^2_2,
\end{equation}

\noindent where $R_h=ar$ is the area radius ($a$ is the scale
factor), $h_{ij}= \textrm{diag}(-1, a^2(t))$  and
$d{\Omega}^2_2=d\theta^2+\sin^2\theta d\phi^2$ is the metric on
the unit 2-sphere. Here $i$, $j$ can take values $0$ and $1$, such
that $x^0=t$, $x^1=r$.

\noindent For a viable $f(R)$ gravity theory, if we take
$f(R)=R+F(R)$, then the modified Friedmann equations in the above
space time can be written as \cite{Wu:2012ia},

\begin{equation}\label{eqn4}
H^2=\frac{8\pi G}{3}\rho_t,
\end{equation}

\begin{equation}\label{eqn5}
\dot{H}=-4\pi G(\rho_t+p_t),
\end{equation}

\noindent where $\rho_t=\rho+\rho_e$ and $p_t=p+p_e$. Here
$\rho_e$ and $p_e$ represents the effective energy density and
effective pressure due to the curvature contribution and they are
given by

\begin{equation}\label{eqn6}
\rho_e=-\frac{1}{2}\Big(F-RF_1+6H\dot{F}_1+6F_1 H^2\Big),
\end{equation}

\begin{equation}\label{eqn7}
\rho_e+p_e=\Big(\ddot{F}_1-H\dot{F}_1+2\dot{H}F_1\Big),
\end{equation}

\noindent where $F_1=\frac{dF}{dR}$, $R=6(\dot{H}+2H^2)$ and from
here onwards we are choosing $8\pi=1$, and $G=1$.

\noindent The energy conservation equations are given by

\begin{equation}\label{eqn9}
  \dot{\rho} + 3H(\rho + p) = 0, \hspace{2cm} \dot{\rho_t} + 3H(\rho_t + p_t) = 0.
\end{equation}

\noindent So the effective pressure and energy density also
satisfies the conservation equation

\begin{equation}\label{eqn11}
  \dot{\rho_e} + 3H(\rho_e + p_e) = 0.
\end{equation}

The above equations will be used in deriving  calculating time
derivatives of  modified entropies in four  alternative
Bekenstein-Hawking formulations given in Sec.IV.


\section{Basic concepts of gravitational thermodynamics}

\noindent In this section we shall study some basic features of thermodynamics. The radius of event horizon $R_E$ is given by

\begin{equation} \label{eqn12}
R_E = a(t)\int^{\infty}_{t}\frac{dt'}{a(t')},
\end{equation}

\noindent and

\begin{equation}\label{eqn13}
\dot{R}_E= HR_E-1.
\end{equation}

\noindent On the other hand radius of apparent horizon in a flat
universe is written as

\begin{equation}\label{eqn14}
R_A=\frac{1}{H}.
\end{equation}

\noindent We know that  the total entropy of an isolated
macroscopic physical system can not decrease (GSLT), $i.e.$
$\dot{S}_T \geq0$, where $S_T$ is total entropy which is the sum
of horizon entropy ($S_h$) and the entropy of the fluid bounded by
the horizon( $S_f$). Also such a system evolves towards TE, a
state having maximum entropy. For universe filled with a fluid and
bounded by event horizon, the validity of GSLT and TE can be
verified using the inequalities given below
\cite{Pavon:2012pt,Mocioiu:2011zz}
\begin{equation}\label{eqn15}
\dot{S}_T\geq0 \qquad \rm (for\ GSLT)
\end{equation}

\noindent and

\begin{equation}\label{eqn16}
\ddot{S}_T<0 \qquad \rm (for\ TE)
\end{equation}

\noindent $S_{f}$ can be calculated from Gibb's equation
\cite{Mazumder:2009zz, Mazumder:2009zza, Mazumder:2010fi,
Chakraborty:2010jx, Chakraborty:2010hr, Izquierdo:2005ku,
Mitra:2015wba, Mitra:2015nba, Dutta:2010zza}

\begin{equation}\label{eqn17}
 {T_f}dS_{f} = dE_f + p dV_E,
\end{equation}
\noindent where $E_f(=\rho V_E)$ is the energy flow across the horizon, $V_E(=\frac{4}{3}\pi R^3_E)$ is the volume of the fluid and $T_f$ is the temperature of the fluid. In the present work as we consider only event horizon, so we assumed the temperature of the fluid is same as the temperature of the horizon, $i.e.$ $T_f=T_E$. Hence the time derivative of fluid entropy is given by

\begin{equation}\label{eqnG}
\dot{S}_f=\frac{1}{2T_f}R^2_E(\rho+p)(\dot{R}_E-HR_E)
\end{equation}

\noindent Using above equation and entropy of the horizon, total
time derivatives in various alternative  Bekenstein-Hawking
formulation can be calculated. In what follows we list four
different temperatures and entropies for examining GSLT and TE in
modified Bekenstein formulation in f(R) gravity.


\section{Modified Bekenstein-Hawking formulation}

\noindent In this section we  list the following four alternative
Bekenstein formulation for possible modification of entropy and
temperature in $f(R)$ gravity. The modifications are carried out
in a manner so that Clausius relation holds on the event horizon.
\vspace{0.1cm}

 \noindent {\bf Case I}: $S_{E}= \frac{R^2_E}{8}$ and $T_E= \frac{4\alpha
 R_E}{R^2_A}$\\

  \noindent {\bf Case II}: $S_{E}= \frac{A f_R}{4}$ and
  $T_E=4{d_1}H$\\

  \noindent {\bf Case III}: $S_{E}= \frac{R^2_E}{8} $ and
  $T_E= \frac{4R_E}{R^2_A}$\\

  \noindent {\bf Case IV}: $S_{E}= \frac{A_E}{4}-\frac{1}{16}\int\Big(\frac{R^2_A R_E}{1-\epsilon}\Big)\Big(\frac{HR_E+1}{HR_E-1}\Big)(\rho_e+p_e)dR_E $ and
  $T_E=4\Big(\frac{R_E}{R_A}\Big)^2\Big(\frac{1-\frac{\dot{R}_A}{2HR_A}}{R_E}\Big), $\\

\noindent  Each case gives  an alternate formulation of Bekenstein
system and in each case  $T_E$ and $S_E$ denote the  temperature
and entropy of the event horizon respectively.  In what follows we
give the motivation and background of each case:

\noindent {\bf Case I}: It is well known that $T_E=\frac{1}{2\pi
R_E}$ and $S_E=\frac{A}{4}$ are called Hawking temperature and
Bekenstein entropy \cite{Hawking:1974sw, Bekenstein:1973ur}.
Recently by generalising Hawking temperature, Chakraborty in ref
\cite{Chakraborty:2012eu} have shown validity of GSLT and TE on
event horizon in Einstein's gravity.

\noindent The generalised Hawking temperature is given by

\begin{equation}\label{eqn19}
T_E=\frac{4\alpha R_E}{R^2_A},
\end{equation}

\noindent where $\alpha=\frac{\frac{v_A}{R_A}}{\frac{v_E}{R_E}}$,
$v_A=\dot{R}_A, v_E=\dot{R}_E$ and for this value of $\alpha$
first law of thermodynamics (FLT) holds on the event horizon
\cite{Chakraborty:2012eu}. In this case we do not change
Bekenstein entropy  given by $ S_E=\frac{R^2_E}{8}$,  but we
change the Hawking temperature to generalised Hawking temperature.

\noindent {\bf Case II}:  The temperature
$T=-\frac{H}{2\pi}(1+\frac{\dot{H}}{2H^2})$ is known as
Hayward-Kodama temperature \cite{Moradpour:2015vpa}.  To avoid the
negative value of the temperature, it is  redefined  and is
written as $T\simeq\frac{H}{2\pi}$ (often called Cai-Kim
temperature \cite{Cai:2005ra}). In general Cai-Kim temperature is
written as $T=\frac{d_1 H}{2\pi}$, where $d_1$ is a real constant
and it shows deviations from Gibbons-Hawking temperature. For
de-Sitter space, we have $d_1=1$. So in this case the event
horizon temperature is taken as Cai-Kim temperature $i.e.$

\begin{equation}\label{eqn23}
T_E=\frac{d_1 H}{2\pi}=4{d_1}H
\end{equation}
\noindent The  Bekenstein entropy in this case is modified because
of  $f(R)$ gravity as \cite{Wald:1993nt, Cognola:2005de}

\begin{equation}\label{eqn24}
S_E=\frac{A f_R}{4}=\frac{A F_2}{4},
\end{equation}

\noindent with $F_2=f_R$ and $A=\frac{R^2_E}{2}$ is the area of the horizon.

\noindent {\bf Case III}: Similar to the apparent horizon the
surface gravity on event horizon can be defined as
\cite{Saha:2012kj} $ \kappa_E=-\frac{1}{2}\frac{\partial
\chi}{\partial R}|_{R=R_E}=\frac{R_E}{R^2_A} $, so the Hawking
temperature can be modified \cite{Chakraborty:2012cw} as $
T^m_E=\frac{\parallel\kappa_E\parallel}{2\pi}=\frac{4R_E}{R^2_A} $
and the above temperature is known as modified Hawking
temperature. The validity of first law of thermodynamics on the
event horizon has been proved using this temperature in standard
gravity in Ref. \cite{Chakraborty:2012cw}

\noindent {\bf Case IV}: In this case, entropy is evaluated from
the validity of unified first law of thermodynamics which was
introduced by Hayward \cite{Hayward:1994bu, Hayward:1997jp,
Hayward:2004dv}. For event horizon if $d\xi^{\pm}_E=dt\mp adr$ is
the one form orthogonal to the surface of the event, then the
tangent vector $\xi_E$ is given by \cite{Mitra:2015jqa}

\begin{equation}\label{ehr}
\xi_E=\frac{\partial}{\partial t}-\frac{1}{a}\frac{\partial}{\partial r}.
\end{equation}

\noindent Now projecting the unified first law along $\xi_E$, one
can write first law of thermodynamics for event horizon as
\cite{Cai:2006rs, Akbar:2006kj, Cai:2007bh}

\begin{equation}\label{ufl}
\langle dE,\xi_E \rangle= \kappa_E\langle dA,\xi_E \rangle +\langle WdV,\xi_E\rangle,
\end{equation}

\noindent and consequently the modified entropy on the event horizon can be evaluated as

\begin{equation}\label{eqn31}
S_E=\frac{A_E}{4}-\frac{1}{16}\int\Big(\frac{R^2_A
R_E}{1-\epsilon}\Big)\Big(\frac{HR_E+1}{HR_E-1}\Big)(\rho_e+p_e)dR_E
\end{equation}

 \noindent where  surface gravity $\kappa_E$ is defined as
 $\kappa_E=-\Big(\frac{R_E}{R_A}\Big)\Big(\frac{1-\epsilon}{R_E}\Big)$
 and $\epsilon=\frac{\dot{R_A}}{2HR_A}$. The entropy expression
 shows the entropy of the event horizon  actually differs from
 Bekenstein entropy by a correction term. Using the above from of surface gravity, one gets extended Hawking
 temperature as \cite{ Mitra:2015nba}
\begin{equation}\label{eht}
T_E=\frac{\parallel\kappa_E\parallel}{2\pi}=4\Big(\frac{R_E}{R_A}\Big)^2\Big(\frac{1-\frac{\dot{R}_A}{2HR_A}}{R_E}\Big),
\end{equation}
 Further, using this form
 of modified entropy and extended Hawking temperature validity of GSLT and
TE have been examined in  various gravity theories in ref
\cite{Mitra:2015wba}. For brevity,  total  variation of entropies
in each case  have been given in appendix. In what follows in
following subsection we shall perform thermodynamic analysis
considering  the universe filled with HDE and make a comparative
study of above four cases in the context of GSLT and TE. It may be
noted that recently this type of analysis have been extensively
done in various gravity theories
\cite{Mitra:2015nba,Mitra:2015wba}


\section{Thermodynamic Analysis for universe filled with holographic dark energy }

\noindent In this section we shall perform thermodynamical
analysis and compare the above four cases for GSLT and equilibrium
thermodynamics. We consider the universe filled with holographic
dark energy interacting with dark matter (DM) in the form of dust.
So its total energy density is $\rho=\rho_m+\rho_d$, where
$\rho_m$ and $\rho_d$ are the energy densities of DM and DE
respectively. As we consider the interaction between DM and DE, so
$\rho_m$ and $\rho_d$ satisfies the following conservation
equations

\begin{equation}\label{eqn35}
\dot{\rho_m} + 3H\rho_m = Q,
\end{equation}

\begin{equation}\label{eqn36}
\dot{\rho_d} + 3H\rho_d(1 + \omega_d) = -Q,
\end{equation}

\noindent where $\omega_d$ is variable equation of state parameter
of DE and $Q=3Hb^2\rho$ is interaction term, with $b^2$ as
coupling parameter.  Also $\omega_d$ satisfies following equation
\cite{Saha:2012kj, Wang:2005jx}

\begin{equation}\label{eqn37}
\omega_d=-\frac{1}{3}-\frac{2\sqrt{\Omega_d}}{3c}-\frac{b^2}{\Omega_d},
\end{equation}

\noindent where $c$ is a dimensionless constant and the density parameter is given by

\begin{equation}\label{eqn38}
{\Omega}^{\prime}_d=\Omega_d\Big[(1-\Omega_d)\Big(1+\frac{2\sqrt{\Omega_d}}{c}\Big)-3b^2\Big],
\end{equation}

\noindent where $\prime=\frac{\partial}{\partial x}$, $x={\rm ln}$ $a$. The velocities of the apparent$(v_A)$ and event$(v_E)$ horizons can be written as

\begin{equation}\label{eqn39}
v_A=\frac{3}{2}\Big[(1-b^2)-\frac{\Omega_d}{3}\Big(1+\frac{2\sqrt{\Omega_d}}{c}\Big)\Big],
\end{equation}

\noindent and

\begin{equation}\label{eqn40}
v_E=\Big(\frac{c}{\sqrt{\Omega_d}-1}\Big).
\end{equation}

\begin{table}[h]
\caption{Planck Data Sets}\label{table1}
\begin{center}
 \begin{tabular}{|c|c|c|c|}
 \hline
 Sl. No. & Data Sets & $c$ & $\Omega_d$ \\
\hline
\hline
 1 & Planck+CMB+SNLS3+lensing & $0.603$ & $0.699$ \\
 \hline
 2 & Planck+CMB+Union $2.1$+lensing & $0.645$ & $0.679$ \\
\hline
 3 & Planck+CMB+BAO+HST+lensing & $0.495$ & $0.745$ \\
 \hline
\end{tabular}
\end{center}
\end{table}

\noindent For graphical representation three data sets have been
used from Table I \cite{Mitra:2015jha, Li:2013dha, Saha:2014uwa,
Pan:2014afa, Saha:2014jja}. From observation, it is found that
Planck data are more accurate than Wilkinson Microwave Anisotropy
Probe(WMAP)-9 data. This accuracy can be increased more if we take
External Astronomical data sets(EADS)  and lensing data into
account. Common EADS include the Baryonic Acoustic Oscillation
(BAO) measurements from 6dFGS+SDSS+DR7(R)+BOSS DR9, Estimation of
Hubble constant  from Hubble Space Telescope (HST) and supernova
data sets SNLS3 together with Union $2.1$.

\noindent Using these data sets graphs of GSLT$(\dot{S}_T)$ and
TE$(\ddot{S}_T)$ have been plotted  each case in FIGS.$1-6$,
considering $H=1$, $R_E=\frac{c}{H\sqrt{\Omega_d}}$ and $d_1=1$.
(See appendix for expressions of  total variation of entropies in
each case)
\newpage
\begin{table}[h]
\caption{Graphical Analysis of GSLT and TE when the universe is
dominated by holographic dark energy}
\begin{tabular}{|c|c|p{4cm}|p{5cm}|}
\hline
Data Set & Case & GSLT &TE\\
\hline \hline
1&I& Holds for $b^2>0.118$& Holds for $b^2<0.228$\\
\hline
1&II&Always holds& Holds for $ 0.24<b^2<0.359$\\
\hline
1&III& Always holds &  Holds for $b^2>0.448$\\
\hline
1&IV& Never hold& Holds for $b^2>0.4$\\
\hline \hline
2&I& Holds for $b^2>0.2$& Holds for $b^2<0.3$ or $b^2>0.53$\\
\hline
2&II&Always holds& Never hold\\
\hline
2&III& Always holds & Holds for $b^2<0.2$ or $b^2>0.41$ \\
\hline
2&IV& Never hold& Holds for $b^2>0.4$\\
\hline \hline
3&I& Always holds &Holds for $b^2<0.244$ or $b^2>0.6$\\
\hline
3&II&Always holds& Never hold\\
\hline
3&III& Always holds & Holds for $b^2>0.526$\\
\hline
3&IV& Never hold&Never hold\\
\hline
\end{tabular}
\end{table}

\begin{figure}[h]
\begin{minipage}{0.45\textwidth}
\includegraphics[width=0.85\linewidth]{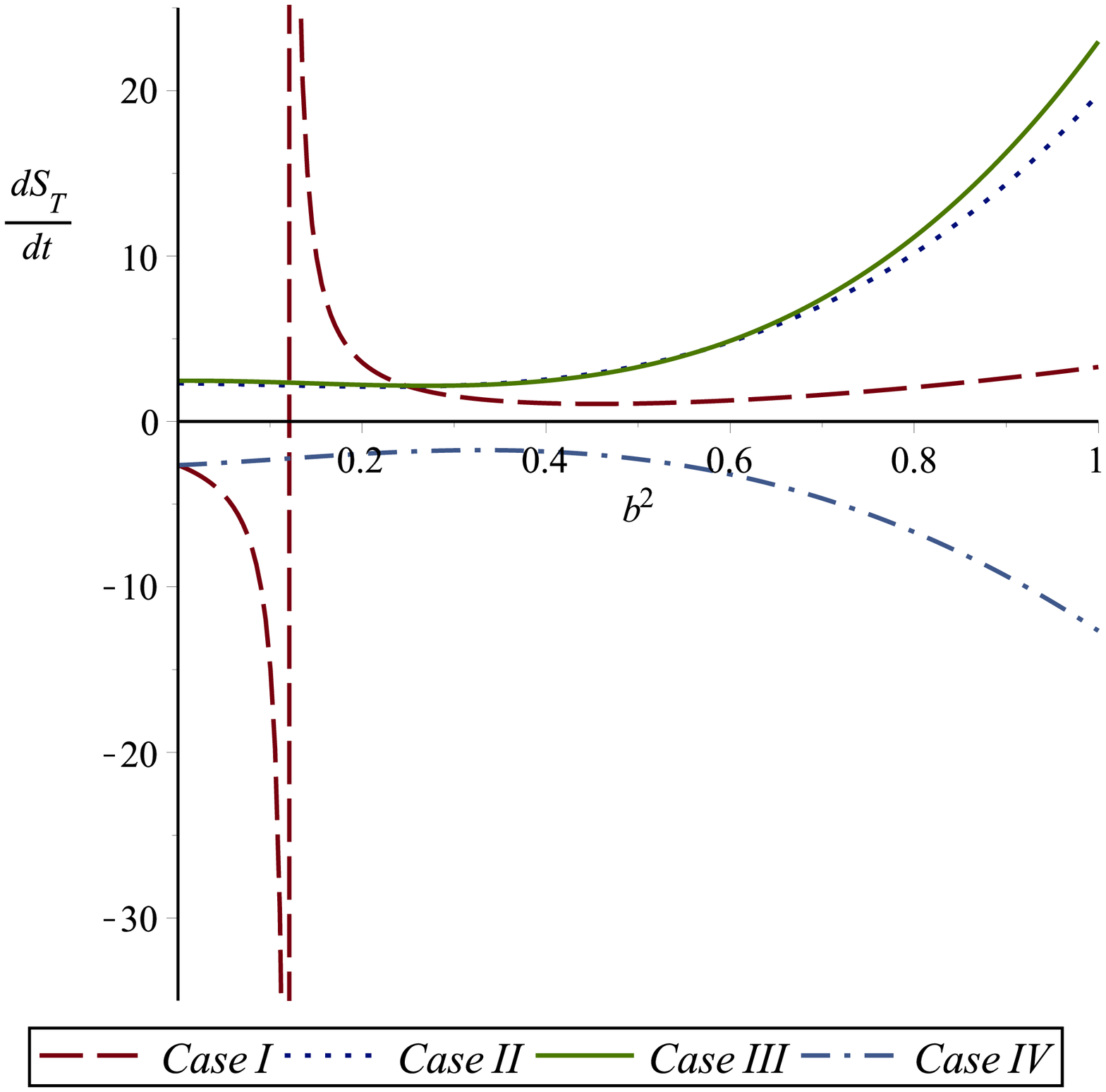}
\caption{The time derivative of the total entropy is plotted
against $b^2$  with $c=0.603$ and $\Omega_d=0.699$.}
\end{minipage}
\begin{minipage}{0.45\textwidth}
\includegraphics[width=0.85\linewidth]{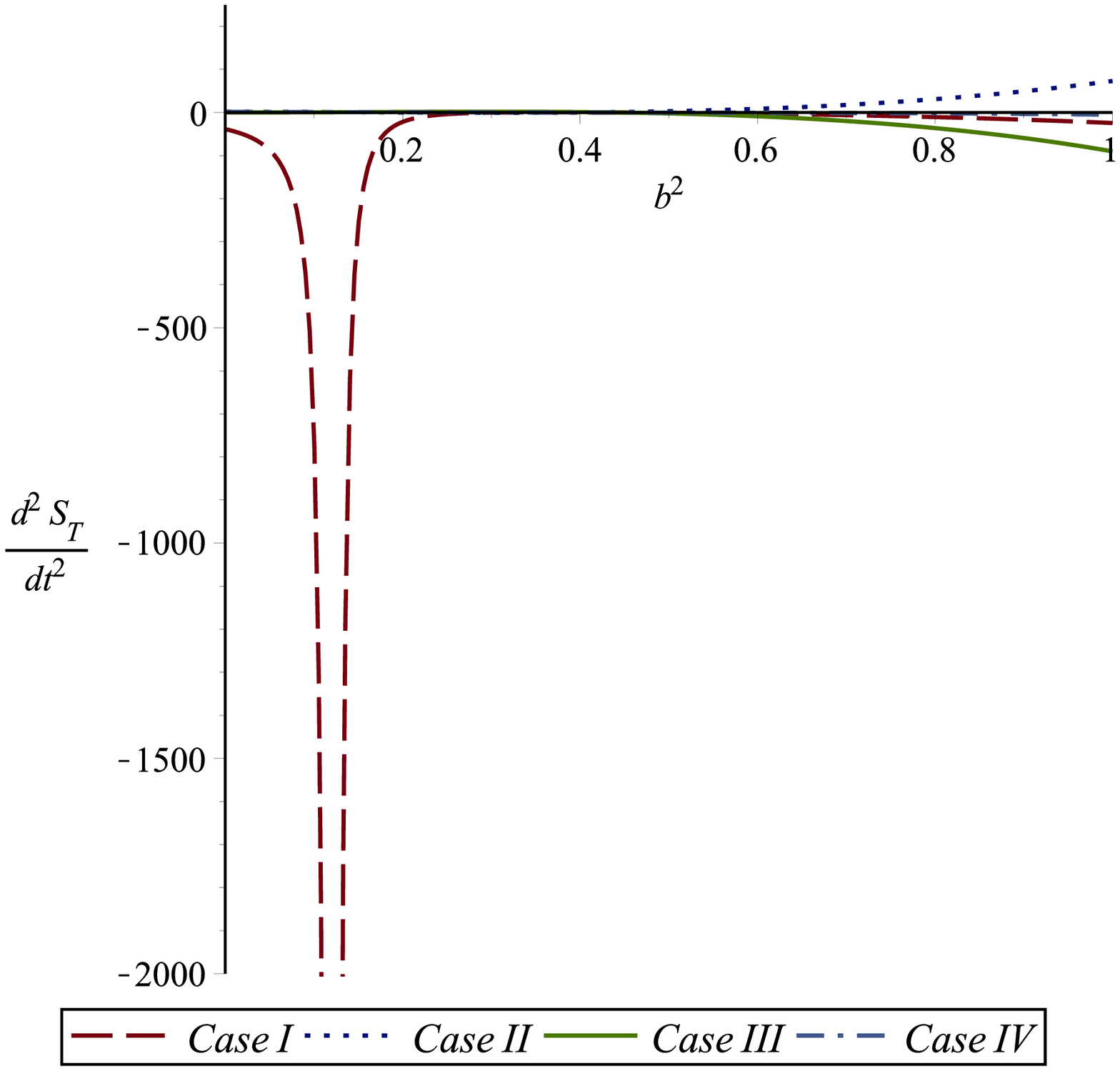}
\caption{The second order time derivative of total entropy is
plotted against $b^2$  with $c=0.603$ and $\Omega_d=0.699$. }
\end{minipage}
\end{figure}


\section{Conclusion}
\noindent This paper deals with the study of thermodynamic
analysis for the universe bounded by event horizon in $f(R)$
gravity theory. In flat FRW model event horizon can only exist in
the accelerating phase of the universe. As from recent observation
the universe is going through an accelerated phase of expansion,
so it is pertinent to consider the universe bounded by event
horizon. Also as dark energy is a possible dominant candidate for
the matter in the universe  and HDE can give effective description
of f(R) gravity. So we have for convenience chosen HDE as the
dominant source of energy. We studied validity of GSLT and TE in
four alternative Bekenstein  formulation. It may be noted that
recently, this type of alternate Bekenstein formulations have been
extensively investigated in various gravity theories. Another
physical motivation to study thermodynamical laws is that if two
cosmological models satisfy equally observational constraints but
one respects thermodynamical laws and other does not, then later
one can be ruled out. So any sensible physical system must satisfy
GSLT and TE. Furthermore, in order to understand the complicated
expressions, the validity of GSLT and TE  are examined
graphically. To have a comparative study of above four different
modified  Bekenstein system, we have examined numerically our
theoretical results with three Planck data sets presented in Table
I.

 The Table-II shows the region where GSLT and TE holds when the
universe is filled with HDE. Our main results can be summarized as
follows from the figures FIG.$1-6$,:

\begin{itemize}
    \item  It is found that GSLT holds
for the Case II and Case III in all three data sets but Case IV
fails in all three data sets.
    \item However, in Case I GSLT holds under
some restriction of $b^2$ in data sets $1$ and $2$ but is not
satisfied in data set $3$.
    \item On the other hand TE holds in all four
cases for first data set but in second data set except Case II,
all other three cases are satisfied under some restriction of
$b^2$.

\item In case of third data sets TE holds only in Case I and Case
III under some restriction of $b^2$ but Case II and Case IV fails.

\end{itemize}

Therefore, from the above comparative study we can conclude that
Case I $i.e.$, Bekenstein entropy with generalised Hawking
temperature and Case III $i.e.$ Bekenstein entropy with modified
Hawking temperature are better compare to other two cases. It may
be noted that in contrast to GR, when Bekenstein entropy is used
then GSLT holds good and TE holds with some restriction but GSLT
is violated when entropy is modified \cite{Chakraborty:2012eu}.
Moreover, by changing the temperature it is shown that GSLT and TE
holds good. So in f(R) gravity the generalised/modified Hawking
temperature has a crucial role in formation of perfect
thermodynamical system. It remains to be seen whether the
generalised/modified Hawking temperature will have same role in
other gravity theories. We leave it for our future work.

\begin{figure}[h]
\begin{minipage}{0.45\textwidth}
\includegraphics[width=0.85\linewidth]{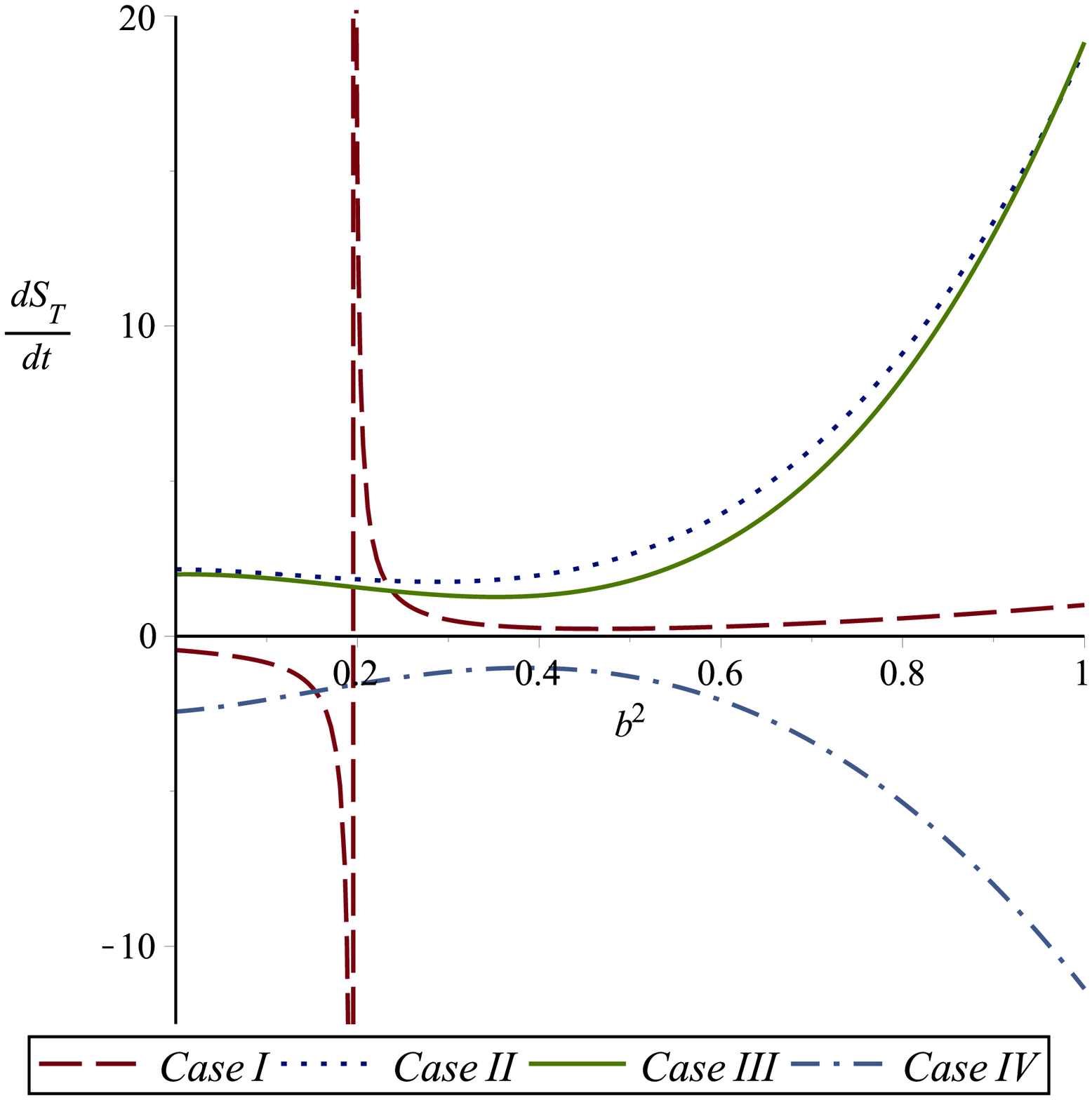}
\caption{The time derivative of the total entropy is plotted
against $b^2$  with $c=0.645$ and $\Omega_d=0.679$.}
\end{minipage}
\begin{minipage}{0.45\textwidth}
\includegraphics[width=0.85\linewidth]{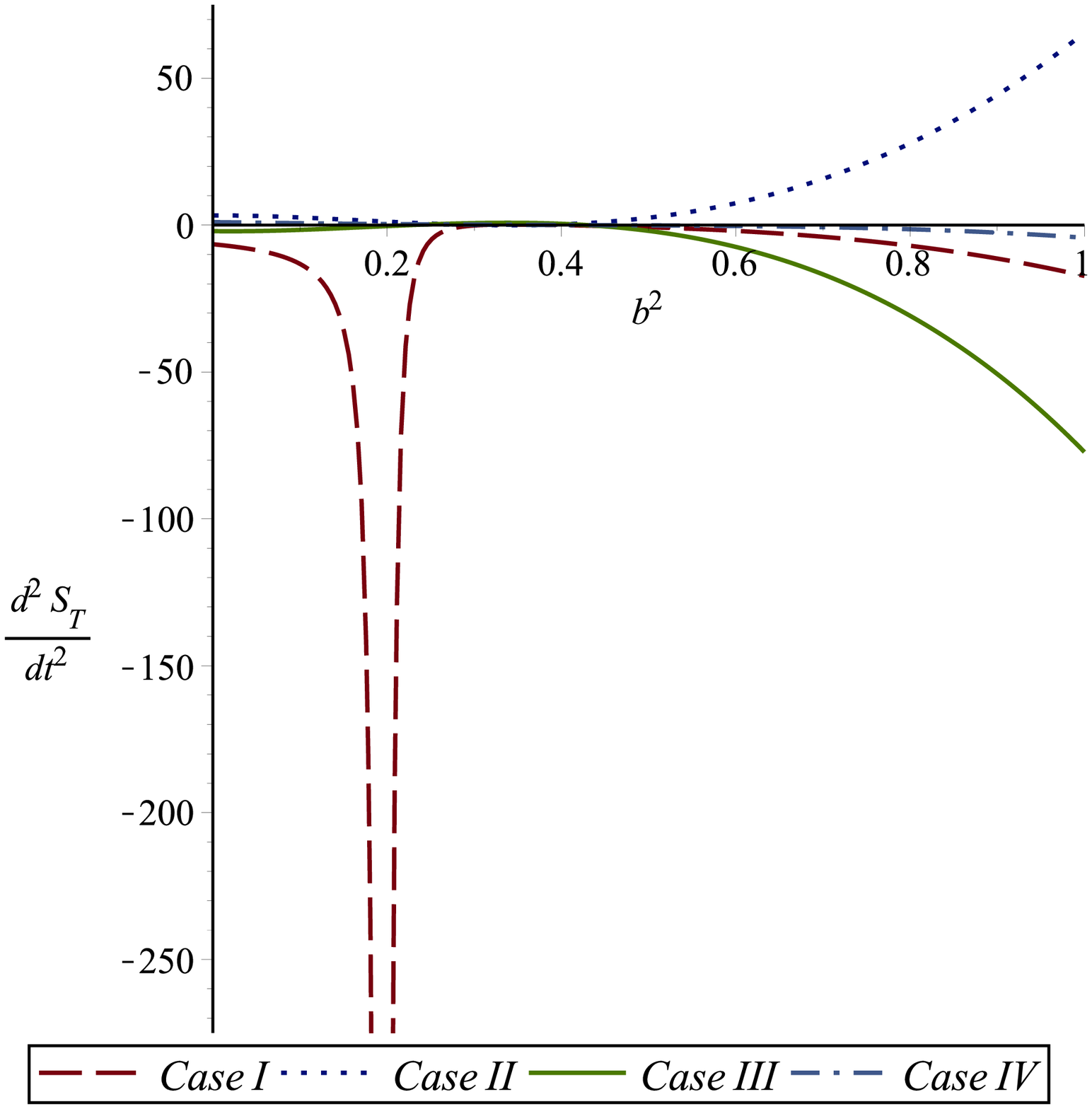}
\caption{The second order time derivative of total entropy is
plotted against $b^2$  with $c=0.645$ and $\Omega_d=0.679$.}
\end{minipage}
\end{figure}

\begin{figure}[h]
\begin{minipage}{0.45\textwidth}
\includegraphics[width=0.85\linewidth]{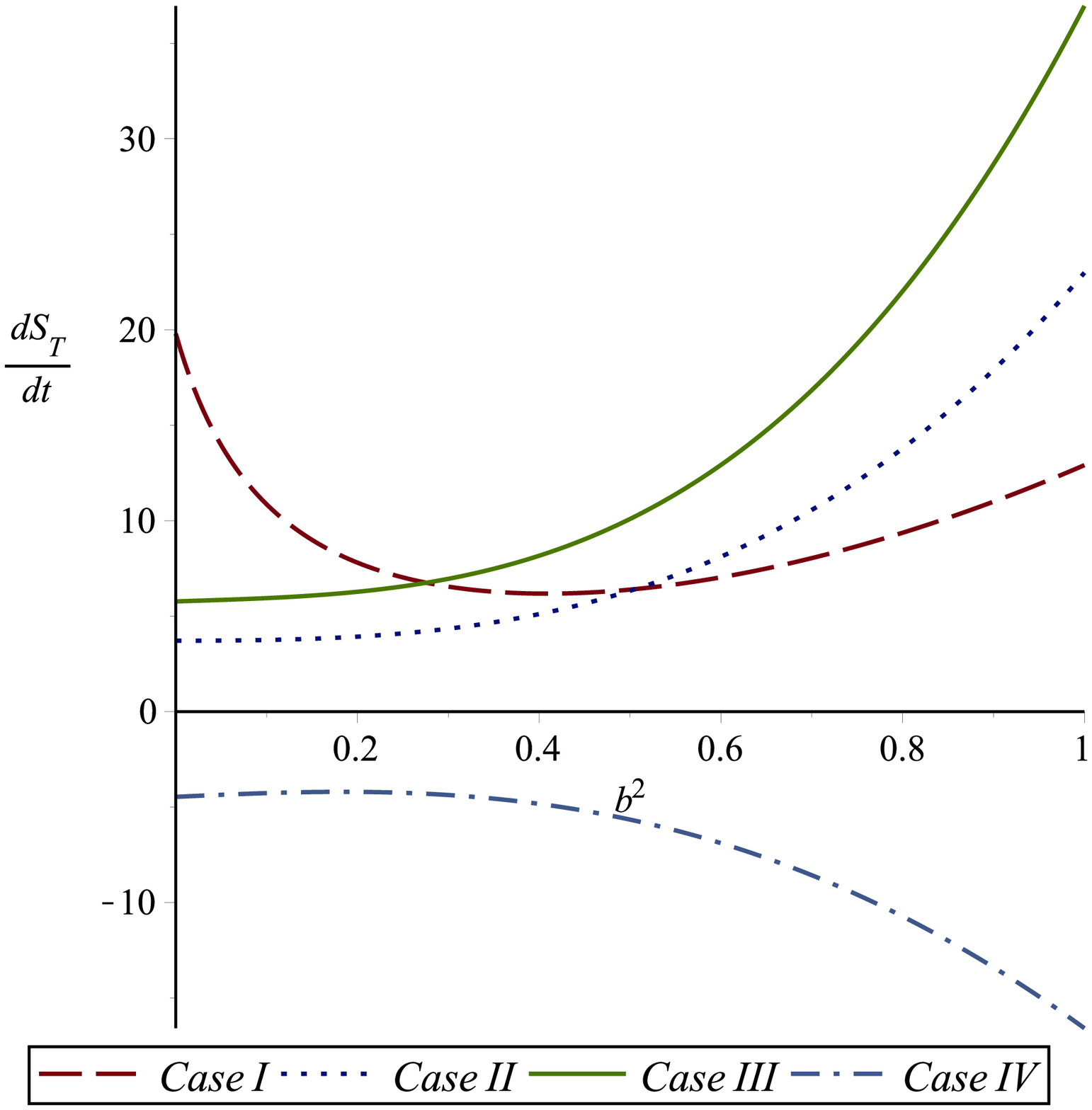}
\caption{The time derivative of the total entropy is plotted
against $b^2$  with $c=0.495$ and $\Omega_d=0.745$.}
\end{minipage}
\begin{minipage}{0.45\textwidth}
\includegraphics[width=0.85\linewidth]{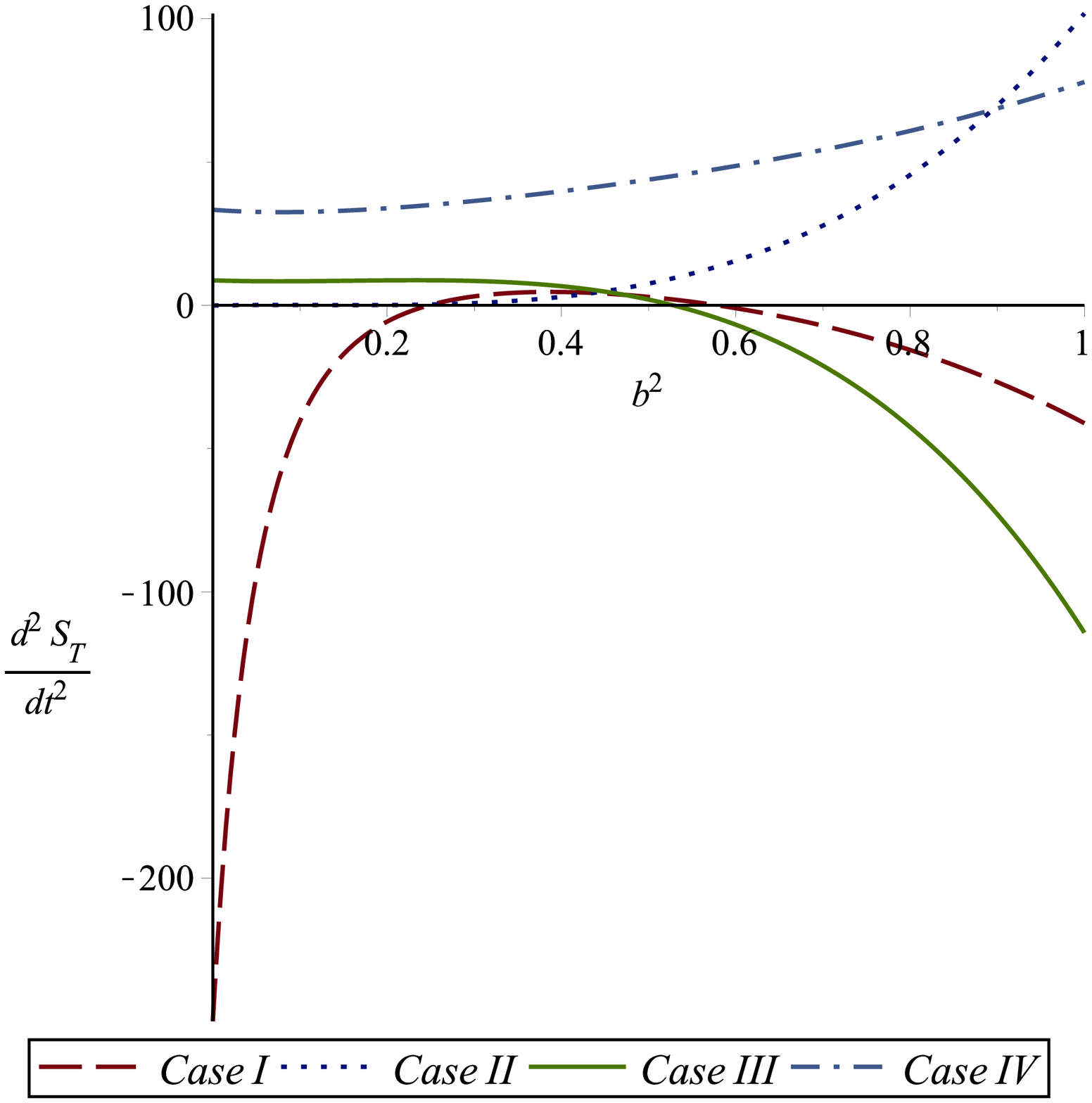}
\caption{The second order time derivative of total entropy is
plotted against$b^2$  with $c=0.495$ and $\Omega_d=0.745$.}
\end{minipage}
\end{figure}

\vspace{3cm}

{\bf Acknowledgement:}\\
The paper is done during a visit to IUCAA, Pune, India. The first
author is thankful to IUCAA for warm hospitality and facility of
doing research works. Further, we would like to thank the referee
for her/his feedback which lead to improvement of the work.

\vspace{4cm}


\appendix
\begin{center}
\textbf{Appendix}\linebreak
\end{center}
In this appendix we will give the expressions of $\dot{S_T}$ and
$\ddot{S_T}$ for four cases of modified Bekenstein-Hawking
formulation. First we note that in each case the the time
variation of $(\rho_e+p_e)$ is given by

\begin{equation}\label{eqn8}
\frac{\partial}{\partial t}
(\rho_e+p_e)=\Big(\dddot{F}_1+\dot{H}\dot{F}_1-H\ddot{F}_1+2\ddot{H}F_1\Big)
\end{equation}
So using Eq.(\ref{eqnG}), Eq.(\ref{eqn8})and definition of $T_E$
and $S_E$ for each case   we get the following expressions :

\noindent {\bf Case I}:

 \noindent Here the first time derivative of total entropy is given by

\begin{equation}\label{eqn21}
\dot{S}_T=\frac{1}{4}\left(R_E v_E-\frac{v_E (\rho +p)}{2v_A
H^3}\right).
\end{equation}

\noindent So the second time derivative of total entropy is given
by

\begin{equation}\label{eqn22}
\begin{split}
\ddot{S}_T=\frac{(R_E f_E+v^2_E)}{4}-\frac{1}{8} \Big[\frac{(\rho
+p)(v_A H^3 f_E-v_E f_A H^3-3v_A v_E H^2\dot{H})}{(v_A
H^3)^2}\\+\frac{v_A v_E H^3\frac{\partial (\rho +p)}{\partial
t}}{(v_A H^3)^2}\Big],
\end{split}
\end{equation}
\noindent where $f_E=\dot{v}_E$.

\noindent {\bf Case II}:

\noindent In this case first time derivative of total entropy is
given by

\begin{equation}\label{eqn25}
\dot{S}_T=\frac{1}{8}\Big[(2R_E v_E
F_2+R^2_E\dot{F}_2)-\frac{R^2_E(\rho +p)}{d_1 H}\Big].
\end{equation}

\noindent The second time derivative of total entropy is

\begin{equation}\label{eqn26}
\begin{split}
\ddot{S}_T=\frac{1}{8}\Big[(2v^2_E F_2+2R_E f_E F_2+4R_E
v_E\dot{F}_2+R^2_E\ddot{F}_2)+\frac{1}{d_1 H^2}\Big(R_E(\rho
+p)(2Hv_E-\dot{H}R_E)\\+HR^2_E \frac{\partial (\rho +p)}{\partial
t}\Big)\Big].
\end{split}
\end{equation}

\noindent {\bf Case III}:

\noindent The first time derivative of total entropy is given by

\begin{equation}\label{eqn29}
\dot{S}_T=\frac{1}{4}\left(R_E v_E-\frac{R_E(\rho
+p)}{2H^2}\right),
\end{equation}

\noindent and second time derivative of total entropy is

\begin{equation}\label{eqn30}
\ddot{S}_T=\frac{(R_E f_E
+v^2_E)}{4}-\frac{1}{8H^3}\Big[(Hv_E-2\dot{H}R_E)(\rho
+p)+HR_E\frac{\partial (\rho +p)}{\partial t} \Big]
\end{equation}

\noindent {\bf Case IV}:

\noindent  The first derivative of total entropy is given by

\begin{equation}\label{eqn33}
\dot{S}_T=\frac{R_E v_E}{4}-\frac{R^2_A
R_E}{2(2-v_A)}\Big[\Big(\frac{v_E+2}{4}\Big)(\ddot{F}_1-H\dot{F}_1+2\dot{H}F_1)+v_A
H^2\Big],
\end{equation}

\noindent and the second derivative of total entropy is

\begin{equation}\label{eqn34}
\begin{split}
\ddot{S}_T=\frac{R_E f_E}{4}\Big[1-\frac{R^2_A
(\ddot{F}_1-H\dot{F}_1+2\dot{H}F_1)}{2(2-v_A)}\Big]+\frac{v^2_E}{4}-\frac{R^2_A
R_E}{8(2-v_A)}\Big[(v_E+2)\{(2\frac{v_A}{R_A}+\frac{v_E}{R_E}+\frac{f_A}{2-v_A})\\(\ddot{F}_1-H\dot{F}_1+2\dot{H}F_1)+(\dddot{F}_1+\dot{H}\dot{F}_1-H\ddot{F}_1+2\ddot{H}F_1)\}+4v_A
H^2\Big[\frac{v_E}{R_E}+\frac{2f_A}{v_A(2-v_A)}\Big]\Big].
\end{split}
\end{equation}


\end{document}